\documentclass[authorversion,nonacm]{acmart}

\usepackage{algorithm}
\usepackage{algorithmic}

\setcopyright{none}
\acmDOI{none}

\acmBooktitle{none}
\acmPrice{none}
\acmISBN{none}

\begin{document}

\title{OpenKinoAI: An Open Source Framework for Intelligent Cinematography and Editing of Live Performances}


\author{Rémi Ronfard}
\email{remi.ronfard@inria.fr}
\orcid{1234-5678-9012}
\affiliation{%
  \institution{Univ. Grenoble Alpes, Inria, LJK, CNRS, Grenoble INP}
}

\author{Rémi Colin de Verdière}
\email{rdc@inria.fr}
\affiliation
{
\institution{Univ. Grenoble Alpes, Inria, LJK, CNRS, Grenoble INP}
}

\renewcommand{\shortauthors}{Ronfard  et al.}

\begin{abstract}
OpenKinoAI is an open source framework for post-production of ultra high definition video which makes it possible to emulate professional multiclip editing techniques for the case of single camera recordings. OpenKinoAI includes tools for uploading raw video footage of live performances on a remote web server, detecting, tracking and recognizing the performers in the original material, reframing the raw video into a large choice of cinematographic rushes, editing the rushes into movies, and annotating rushes and movies for documentation purposes. OpenKinoAI is made available to promote research in multiclip video editing of ultra high definition video,
and to allow performing artists and companies to use this research for archiving, documenting and sharing their work online in an innovative fashion. 
\end{abstract}

\begin{teaserfigure}
\begin{center}
   \includegraphics[width=\linewidth]{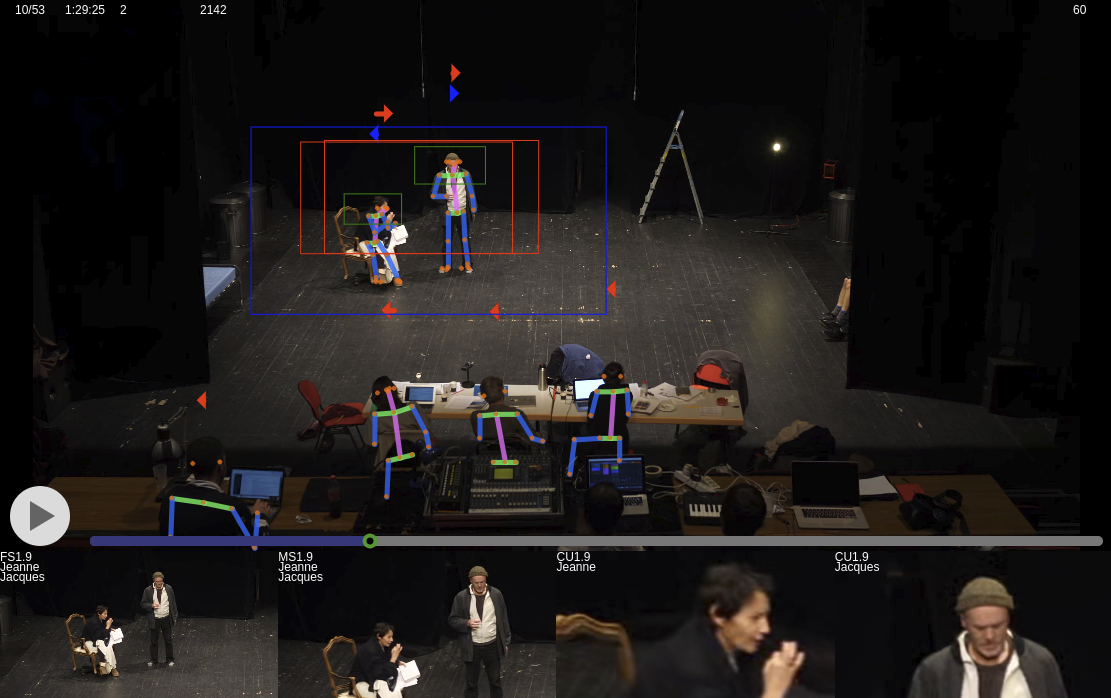}
\end{center}
   \caption{Kinoai workflow. Top: The scene is recorded with a fixed camera in ultra high definition. Bottom: We generate candidate shots with different compositions on the server side and allow remote users to edit them on the client side.}
\label{fig_episode_three}
\end{teaserfigure}


\maketitle
\section{Introduction}
Adapting a stage performance to screen is a difficult task which presents many challenges even to expert human cinematographers and film editors\cite{Gandhi2014,Ronfard2015,Kumar2017}. We propose the OpenKinoAI framework as a toolset for making AI-based video production accessible to theater professionals and scholars.

Previous work in AI-based video production has  investigated video lectures \cite{Zhang2005}, personal videos \cite{Gleicher2008}, talk shows \cite{Wright2020} and sports \cite {Chen2018}. OpenKinoAI is the first plateform directly addressing the challenges of artistic live performances such as theater, opera and dance. OpenKinoAI is an open source framework for ingesting video recordings of live performances, breaking them down into meaningful cinematographic rushes and editing them into movies. 

Our framework provides support for the various tasks involved in video production, including the processing and handling of ultra high definition video and their metadata between a web server and a web client, the synchronized display of the generated movies, and the necessary graphical user interfaces for handling and manipulating multiple timelines. We make the simplifying assumption that the generated movies maintain the temporal continuity of the recorded performances. As a result, we only provide support for emulating {\em live multi-camera editing} in post-production, rather than general non-linear three-point editing, which is left for future work.

\begin{figure}[t]
\begin{center}
   \includegraphics[width=\linewidth]{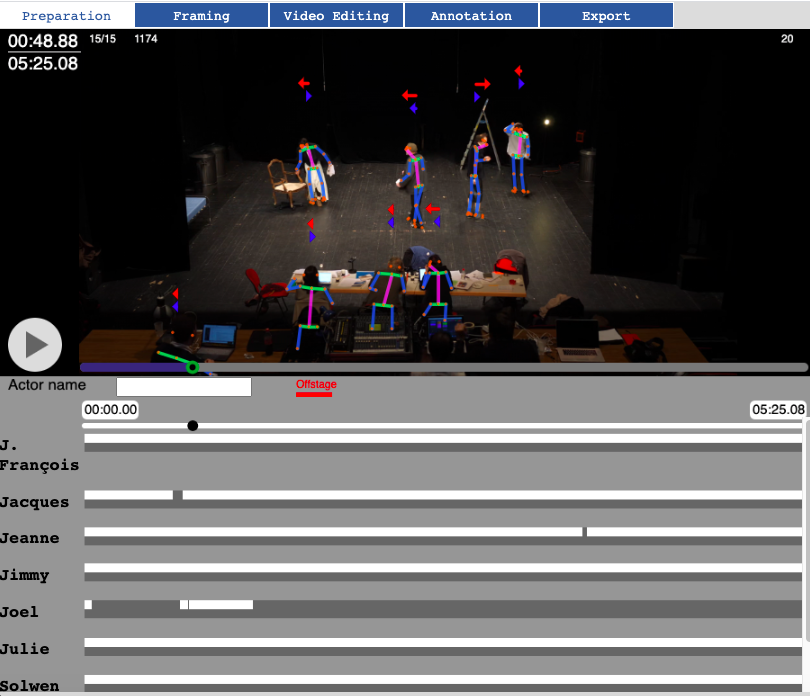}
\end{center}
   \caption{Preparation view. OpenKinoAI provides tools for labeling tracklets from OpenPose detections and assigning them to timelines.}
\label{preparation}
\end{figure}

\section{Software architecture}
OpenKinoAI is an open  software environment which implements rush generation methods from previous work \cite{Gandhi2014,Gandhi2015,Ronfard2016} and provides an API for future work in intelligent film editing of those rushes. Our software architecture carefully separates tasks in cinematography, which are implemented on the server side, from tasks in film editing, which are implemented on the client side.  We describe the main functionalities currently available in OpenKinoAI  in terms of video processing (Section \ref{video_processing}), rush generation (Section \ref{rush_generation}), video editing (Section \ref{multicamera_editing}) and annotation (Section \ref{annotation}). We then describe extensive experimental results obtained in the last two years while implementing the system and discuss the limitations and perspectives of the OpenKinoAI software.

\section{Preparation}
\label{video_processing}
We compress the original (4k or 6k) source with HEVC codec, and create three low resolution sources (360p, 540p and 1080p), one audio source and create a dash manifest \cite{Sodagar2011} for web viewing. In order to provide fast and comfortable viewing experience, we choose to split the footage into 10 minute part and process them separately. 

\subsection{Actor detection}
We use the Openpose toolkit to detect a set of 25 keypoints for each actor present on stage  \cite{OpenPose}. We obtain a 2D skeleton for every actor on stage at every frame. The skeleton is based on the dataset Body 25, a point of the skeleton is a 2D coordinates plus a confidence value which express the probability of the detection.

\subsection{Tracklet generation}
Openpose detection gives a set of actors skeletons for each frame but it does not allow to track actor position over time. The goal here is to link the actor detection through the time with an IoU comparaison between frame t and t+1. A tracklet is defined as a sequence of openpose detections. To create the sequence, first we iterate over frame. For each actor detection in the frame, we compute an upper bounding box with the neck, nose, eyes and ears position. Then we compare the intersection over union with the previous frame. We link the indexes of detections which have the best intersection over union ratio.

\subsection{Actor identification}
Once we have generated a small enough number of tracklets,  we need to attach them with the coreesponding actors names. In previous work, we learned appearance models for each actor. In practice, however, actors change costumes quite frequently, either off stage between scenes, or even on stage in the course of a single scene. As a result, the appearance models need to be update quite frequently. Lighting may also change the appearance of actors quite dramatically. As a result, we instead assign each tracklet with an actor manually in the web interface.

\begin{figure}[t]
\begin{center}
   \includegraphics[width=\linewidth]{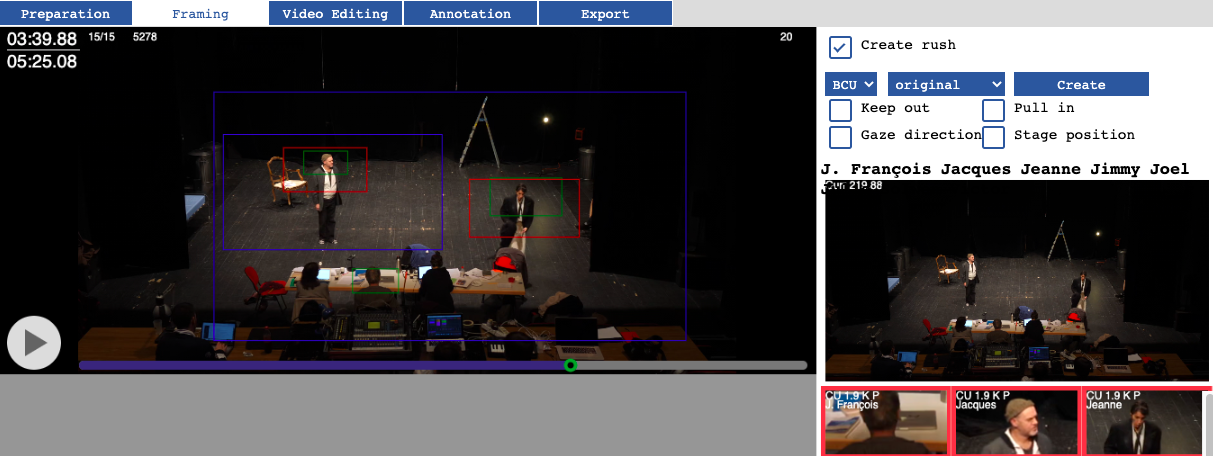}
\end{center}
\caption{Framing view. OpenKinoAI provides tools for choosing shot sizes and compositions and previewing the generated rushes.}
\label{framing}
\end{figure}

\section{Framing}
\label{rush_generation}
OpenKinoAI implements a generic, optimization-based reframing method first described by Gandhi et al. \cite{Gandhi2014}, for automatically reframing high resolution video into a variety of shot types and shot sizes. The new implementation is efficient and robust to large video file sizes, and has been tested on extended  live performances lasting up to two hours in resolution up to $6K$.




\subsection{Computing desired camera frames}
Given a desired shot size and subject matter, we compute a desired camera frame at each time step in the video, 
taking into account stage constraints (S), gaze constraints (G) and composition constraints (Keepout and Pullin). Gandhi et al. take special care of undesired actors who accidentally appear within the camera frame by choosing between two possible strategies. Whenever possible, undesired actors are {\em kept out} of the camera frame with a repulsive force applied to the camera frame during optimization. When undesired actors come too close to the desired actors, this strategy may fail. Instead, undesired actors can be {\em pulled in} to the camera frame in those cases. In OpenKinoAI, we take a slightly different approach where we momentarily change the subject matter and size of the desired shot as needed to pull in or keep out the undesired actors.





\subsection{Best effort optimization}
Given the desired framings at each time step, we then stabilize the camera by solving a large convex otimization problem, as described by Gandhi et al. \cite{Gandhi2014}. For efficiency, this computation is performed in slices of ten seconds and the continuity between slices is enforced with hard constraints. The computation time is fast and independent of the video resolution. 



\begin{figure}[t]
\begin{center}
   \includegraphics[width=\linewidth]{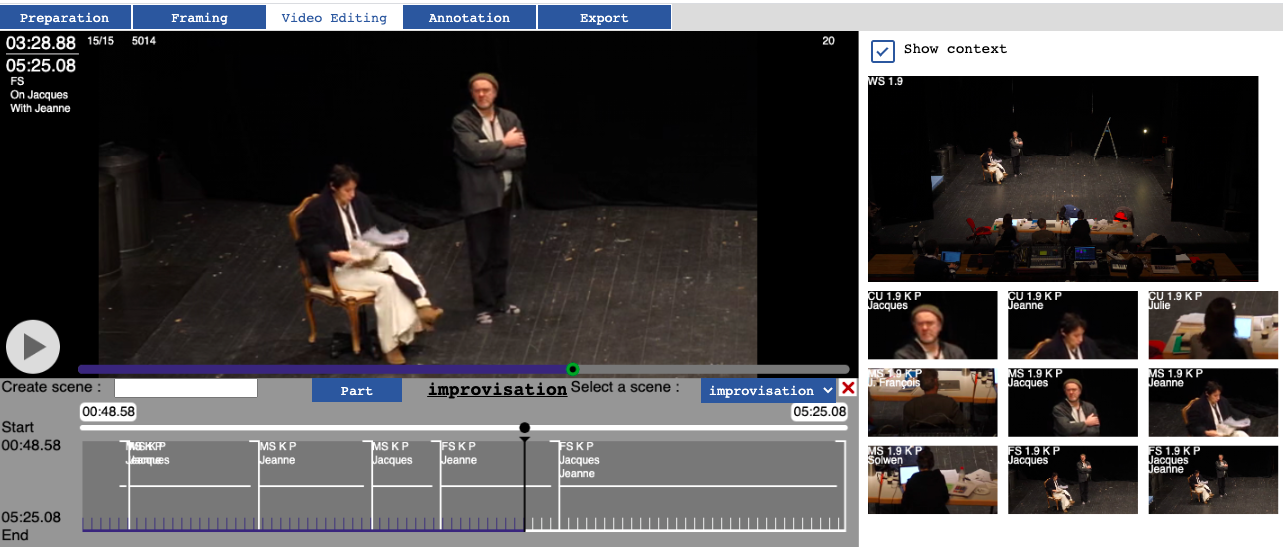}
\end{center}
   \caption{Video editing view. OpenKinoAI provides tools for online  video editing using all available rushes into movie scenes.}
\label{editing}
\end{figure}

\section{Multicamera editing}
\label{multicamera_editing}

\begin{figure}[t]
\begin{center}
   \includegraphics[width=\linewidth]{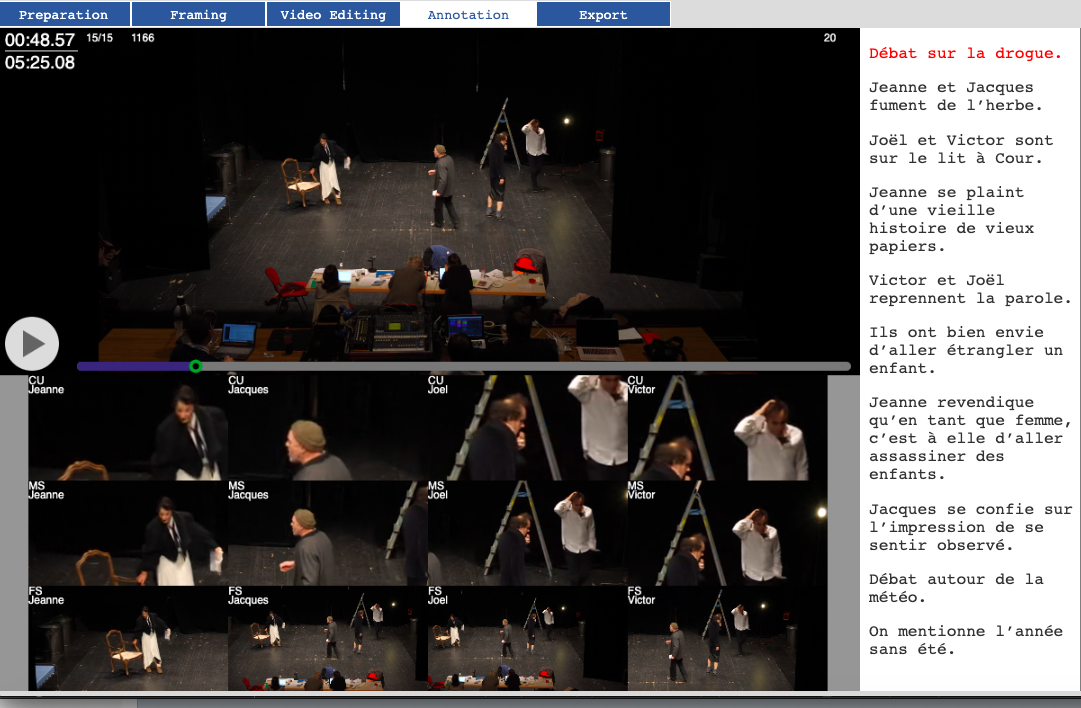}
\end{center}
   \caption{Annotation view. OpenKinoAI provides tools
   for annotating rushes and movie scenes. Annotations
   are displayed as video description subtitles.}
\label{montage}
\end{figure}

OpenKinoAI provides tools for online video editing using all available rushes into movie scenes. The client application is shown in Figure \ref{montage}. Users can drag and drop rushes to a timeline and refine the editing by moving transition between shots interactively. This mode of editing mimics live editing techniques used in real-time multi-camera production \cite{Jacobson2010}. Both the rushes and the edited movie scenes inherit the metadata computed during preparation (i.e. names and screen locations of performing actors, gaze directions and moving directions).

\section{Annotation}
\label{annotation}

In addition to rush generation and multicamera editing, OpenKinoAI provides tools for transcribing and annotating live performances.  Those tools are engineered to make the best possible use of all available shot sizes and compositions. Typically, speech transcription can be facilitated by showing close up shots of the speaking actors, allowing to quickly identify speakers and transcribing their speech even in  cluttered conversation, with loud music and sound effects on stage. Stage directions annotation can be facilitated by showing full shots of the moving actors together with an ensemble shot of the entire stage. The annotation of other scenographic devices (lighting, video projection, stage decoration) can be facilitated by showing a wide shot of the entire stage.



\section{Impact on academia and industry}

\begin{figure}[t]
\begin{center}
   \includegraphics[width=0.85\linewidth]{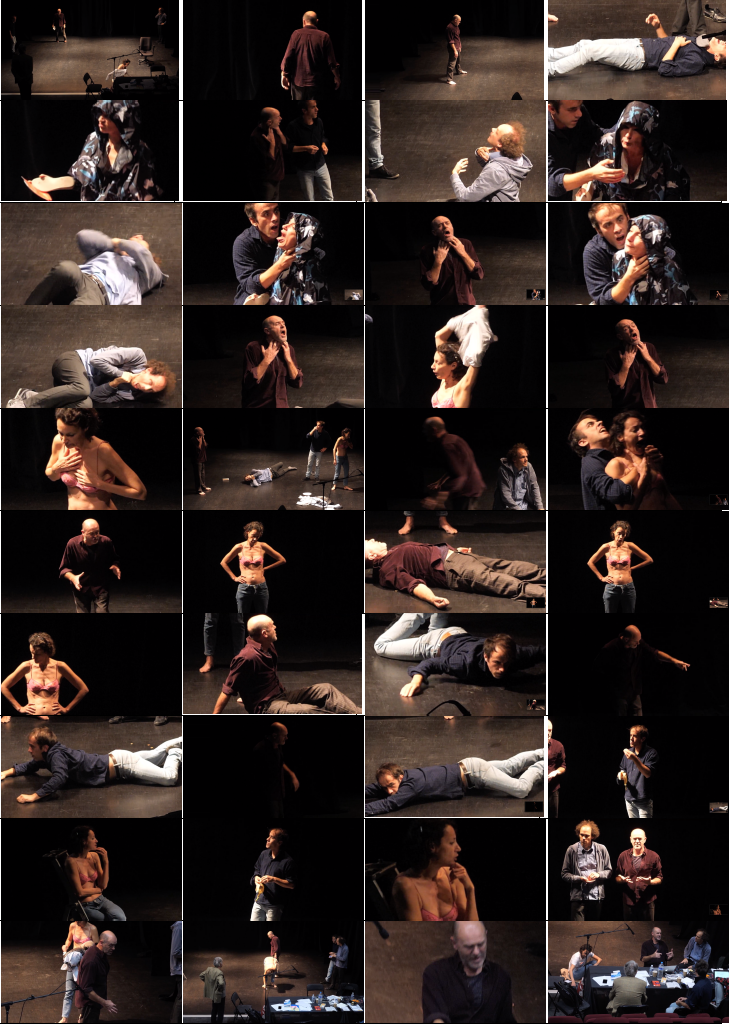}
\end{center}
   \caption{Forty shots from a five-minute movie scene created by a professional film editor using  a single camera recording of a stage adaptation of Mary Shelley's Frankenstein by Jean-François Peyret, Théâtre de Vidy, Lausanne, 2018.}
\label{storyboard}
\end{figure}

OpenKinoAI is an open source software architecture which makes it possible for theatre companies to gain better artistic and technical control over the cinematography and the editing of single-camera video recordings. With ultra high definition cameras now available at reasonable prices, we hope OpenKinoAI will be useful for creating movies of live performances and publishing them online. The OpenKinoAI toolset emulates professional video production tools which were previously not available to a general audience. This is important for theatre companies who can use our system to build memories of their creative work and make them available to the general public at a limited cost. Movies produced in post-production with OpenKinoAI generally reach a higher quality than live streamed videos, at a very similar cost. We hope this can help build a more comprehensive internet archive of world wide live performance movies. 

Starting in October 2020, OpenKinoAi is made available both as an online service \footnote{\url{https://kinoai.inria.fr/}} allowing end-users to freely upload, prepare, reframe, edit and annotate their own footage; and as a software toolkit \footnote{\url{https://gitlab.inria.fr/ronfard/openkinoai}} allowing software developers to configure their own server for non-commercial applications.
   
We hope the software can be useful to other researchers in intelligent cinematography and editing as a generic environment where they can implement their own tools for generating cinematographic rushes and editing them into movies.

\section{Summary and future development}

OpenKinoAI provides all the necessary tools for generating and annotating cinematographic rushes from raw video footage in ultra high definition. OpenKinoAI enforces a strict separation between heavy duty video processing on the server side, and interactive control and immediate feedback on the client side. Directions for future work include support for split-screen composition \cite{Kumar2017}, automatic rough-cut editing \cite{Leake2017} and non-linear three-point editing. OpenKinoAI is distributed in the hope that it can serve as a common test bed and application framework for future research in intelligent cinematography and editing and facilitate their deployment in real life applications.

\begin{acks}
Development of KinoAI was made possible by Inria 
and the Performance Lab at Univ. Grenoble Alpes. 
We thank Jean-François Peyret, Jeanne Balibar, Jacques Bonnaffé, Victor Lenoble and Joel Maillard for allowing us to reproduce images of their rehearsal work. We thank Julie Valéro for numerous discussions and her continuous support and advice in this project. We thank Frédéric Devernay for invaluable contributions to earlier versions of the project. 
\end{acks}

\bibliographystyle{ACM-Reference-Format}
\bibliography{kinoai}


\end{document}